\documentstyle[twocolumn]{jpsj}
\input epsf.tex

\def\runtitle{Charge Kondo effect toward a non-Fermi-liquid fixed point in the orbitally degenerate exchange model}
\def\runauthor{Hiroaki {\sc Kusunose} and Yoshio {\sc Kuramoto}}

\title
{
Charge Kondo effect toward a non-Fermi-liquid fixed point\\
in the orbitally degenerate exchange model
}

\author
{
Hiroaki {\sc Kusunose}\footnote{E-mail: kusu@cmpt01.phys.tohoku.ac.jp}
and Yoshio {\sc Kuramoto}
}

\inst
{
Department of physics, Tohoku University, Sendai, 980-8578
}

\recdate
{
\today
}

\abst
{
We show that a Kondo-type model with an orbital degeneracy has a new
non-Fermi-liquid fixed point.
Near the fixed point the spin degrees of freedom are completely
quenched, and
the residual charge degrees of freedom lead to the multi-channel Kondo
effect.
Anomalous behavior appears in electric and thermal properties, but the
magnetic susceptibility should show the local Fermi-liquid behavior.
The non-Fermi-liquid fixed point becomes unstable against
perturbations breaking the particle-hole symmetry.
We derive these results using the third-order scaling for a
spherically symmetric model with a fictitious spin.
In contrast to the Coqblin-Schrieffer model, the present model
respects different time-reversal properties of multipole operators.
}

\kword
{
orbital degeneracy, time-reversal property, multipole,
multi-channel Kondo effect, non-Fermi liquid, CeB$_6$, LaMnO$_3$
}
\begin{document}
\sloppy
\maketitle

%
%

Orbital dynamics has recently attracted much interest because of new
experimental progress
in heavy-fermion systems such as CeB$_6$\cite{hiroi,nakamura,tayama}
and transition metal oxides such as LaMnO$_3$\cite{ramirez,murakami}.
In such systems, the localized ground multiplet has an orbital
degeneracy $n$ in addition to the Kramers degeneracy.
Thus orbital tensors as well as spin ones constitute
observables.\cite{kugel,schmitt,shiina,fukushima}
The simplest model to describe the combined spin and orbital degrees
of freedom is
the SU($2n$) symmetric spin-orbit
model.\cite{sutherland,affleck,yamashita,li,frischmuth}
In this model, however, the different time-reversal properties of the
spin and orbital tensors are hidden in the excessively high symmetry.

One can naturally ask how the Kondo effect occurs in the presence of
the orbital degeneracy, and how the anisotropy in the combined
spin-orbit space influences the physical property.
In this context we refer to the pioneer work of
Hirst who extended the Coqblin-Schrieffer (CS) model for the realistic
atomic level structure.\cite{hirst}
Pavarini and Andreani recently studied the stability of the local
Fermi liquid in the CS model against the symmetry breaking interaction
found by Hirst, and concluded that it is stable.\cite{pavarini}
In order to include the realistic crystalline-electric-field (CEF)
effect, the present authors considered more general exchange
interaction between $f$ and conduction electrons based on the
point-group analysis, and discussed how the time-reversal property of
operators is reflected in the Kondo effect.\cite{kusunose}
In general, operators describing the spin and/or the orbital degrees
of freedom are classified according to their definite time-reversal
properties.
The operators which are even under the time reversal are called
electric, and those which are odd are called magnetic.
The distinction between magnetic and electric characters should be
important in interpreting and predicting experimental results
including
dilution effects.

In this Letter we propose a new mechanism to realize the
non-Fermi-liquid fixed point by combination of spin and orbital
degrees of freedom.  Near the fixed point
only the spin degrees of freedom are compensated, and the
residual charge degrees of freedom leads to the multi-channel Kondo
effect.
This non-Fermi-liquid fixed point requires the presence of
particle-hole symmetry, and
the anomalous behavior appears only in electric properties.
Namely the magnetic susceptibility behaves as if the system is a local
Fermi liquid.

Let us start with our choice of the basis set and explain their
time-reversal property.
The ground multiplet is written as $|l\sigma\rangle$ ($l=1,2,\cdots
n$, $\sigma=\uparrow,\downarrow$), where $\sigma$ represents the
Kramers pair with arbitrary strong spin-orbit interaction.
The time-reversal operation $\theta$ is defined as
\begin{equation}
\theta|l\sigma\rangle=(-1)^{1/2-\sigma}|l-\sigma\rangle.
\end{equation}
The conventional Pauli matrices $\sigma^\alpha$ acting only on the
spin state are odd
under time reversal:
\begin{equation}
\theta\;\sigma^\alpha\;\theta^{-1}=-\sigma^\alpha,\;\;\;(\alpha=x,y,z).
\end{equation}
We define the orbital operators $\tau^i$ which act only on the orbital
state $l$.
Their time-reversal properties are represented as
\begin{equation}
\theta\;\tau^i\;\theta^{-1}=(\tau^i)^*,\;\;\;(i=1,2,\cdots,n^2-1),
\end{equation}
due to the antilinearity of the time-reversal operation.
Thus, in the Hermitian representation the pure imaginary orbital
operators are odd under time reversal, and have the magnetic
character.
In the simplest case of $n=2$, the operator $\tau^y$ is magnetic and
$\tau^x$ and $\tau^z$ are electric.
In the case of $n=3$, three of eight orbital operators are magnetic
and five are electric.
In the absence of the spin-orbit interaction, magnetic orbital
operators describe the orbital moment of electrons.

In order to facilitate the scaling analysis,
we map the states onto $|m\rangle$ where $m$ represents the
$z$-component of a fictitious spin $j=n-1/2$ (half integer).
An example of mapping for $n=3$ is given by
\begin{eqnarray}
&&|1\uparrow\rangle\to|+5/2\rangle,\;
|2\downarrow\rangle\to|+3/2\rangle,\;
|3\uparrow\rangle\to|+1/2\rangle,\nonumber\\
&&|3\downarrow\rangle\to|-1/2\rangle,\;
|2\uparrow\rangle\to|-3/2\rangle,\;
|1\downarrow\rangle\to|-5/2\rangle.
\end{eqnarray}
We assume that the Hamiltonian has the spherical symmetry in the new
basis set.
Then we introduce the spherical tensor operators\cite{hirst,landau}
which have a simple time-reversal property:
\begin{equation}
\theta\;T^{(p)}_q\;\theta^{-1}=(-1)^p[T^{(p)}_q]^\dagger,
\end{equation}
where each operator represents $2^p$th magnetic (electric) multipole
for odd (even) $p$ with the component $q$.
This property under time reversal merely reflects the odd character of
the first rank operator, the angular momentum.

%
%
The set of the spherical tensors with $p=1,2,\cdots,2j$ together with
the unit matrix completely describes the  $2n\times 2n$ kinds of
transition between local electron states.
Then any spherically symmetric interaction among operators between $f$
and conduction electrons is described in terms of the spherical
tensors as
\begin{equation}
H_{\rm int}=\sum_{p=1}^{2j}J_p\sum_{q=-p}^p
c^\dagger {\sf T}^{(p)}_q c\; |f\rangle [{\sf T}^{(p)}_q]^\dagger
\langle f|,
\label{H_int}
\end{equation}
where $J_p$ is the coupling constant,
${\sf T}^{(p)}_q$ is a matrix representation of the tensor operator
with dimension $2n$
and summation over quantum numbers of each state is implied by the
matrix multiplication.
Here the spherical tensors are orthonormalized as
\begin{equation}
{\rm Tr}({\sf T}^{(p)}_q[{\sf
T}^{(p')}_{q'}]^\dagger)=\delta_{pp'}\delta_{qq'}.
\end{equation}
In general, the angular momentum $j$ of conduction electrons is not
necessarily the same as the localized electron.\cite{multi,cox}
However, we assume the same $j$  for simplicity.
In the case of $J_p=J_{\rm CS}$ for all $p$, one recovers the
Coqblin-Schrieffer model with
the SU($2n$) symmetry.
It should be emphasized that the spherical symmetry SU(2) in
eq.(\ref{H_int}) is adopted just for technical convenience.
Obviously the actual system with the CEF is not invariant under the
infinitesimal rotation.
The simplified model presented above is sufficient for our purpose of
analyzing the consequence of different time-reversal properties
between electric and magnetic degrees of freedom.
Because the time-reversal property of operators is conserved under the
discrete rotation of the point-group, the extension of the model is
straightforward.

A general formula is available for constructing the scaling equation
in terms of the structure constants of the algebra for the related
matrices.\cite{kuramoto,kusunose}
The commutation rule is given by
\begin{equation}
[{\sf T}_\alpha,{\sf
T}_\beta]=\sum_{\gamma}if_{\alpha\beta\gamma}[{\sf T}_\gamma]^\dagger,
\end{equation}
where the Greek $\alpha$, $\beta$ and $\gamma$ are abbreviation of
($p,q$), ($p',q'$) and ($p'',q''$), respectively.
We adopt the completely antisymmetric structure constant which is
given by
\begin{equation}
f_{\alpha\beta\gamma}=2(-1)^{2j-(p+p'+p''+1)/2}S^{(j)}_{pp'p''}
\left(\begin{array}{ccc}
p & p' & p'' \\ q & q' & q''
\end{array}\right),
\end{equation}
where the parenthesis including three columns is the Wigner's $3j$
symbol\cite{landau}.
The symmetric function $S$ with respect to $p$'s is defined as
\begin{equation}
S^{(j)}_{pp'p''}=\sqrt{(2p+1)(2p'+1)(2p''+1)}
\left\{\begin{array}{ccc}
p & p' & p'' \\ j & j & j
\end{array}\right\},
\end{equation}
with $p+p'+p''=$odd.
The bracket including three columns represent the $6j$
symbol\cite{landau}.

Once the commutation rule is given, we can easily obtain the scaling
equations up to third order:
\begin{eqnarray}
&&\frac{\partial}{\partial\xi}J_p=\beta^{(2)}_p+\beta^{(3)}_p\label{sceq1}\\
&&\mbox{\hspace{1cm}}\label{sceq2}
\beta^{(2)}_p=-\frac{2}{2p+1}\sum_{p'p''}J_{p'}J_{p''}(S^{(j)}_{pp'p''})^2,
\\&&\mbox{\hspace{1cm}}\label{sceq3}
\beta^{(3)}_p=\frac{2J_p}{2p+1}\sum_{p'}J_{p'}^2\sum_{p''}(S^{(j)}_{pp'p''})^2,
\end{eqnarray}
where the coupling constants are measured in units of the bare cut-off
energy $D_0$, and $\xi=\ln(D/D_0)$.

Let us simplify the analysis by putting $J_p=J_m$ for odd $p$ and
$J_p=J_e$ for even $p$.
Namely we pay attention only to the difference in time-reversal
property of operators.
Then eqs. (\ref{sceq1})--(\ref{sceq3}) are reduced to
\begin{eqnarray}
&&\frac{\partial}{\partial\xi}J_m=-2(1-J_m)(A_m^{(j)}J_m^2+A_e^{(j)}J_e^2),\label{sc1}\\
&&\frac{\partial}{\partial\xi}J_e=2A_c^{(j)}J_e\left[(J_m^2+J_e^2)-2J_m\right],\label{sc2}
\end{eqnarray}
where $A$'s are defined as
\begin{eqnarray}
&&A_m^{(j)}=\frac{1}{2p+1}\sum_{p'p''}^{\rm
odd}(S^{(j)}_{pp'p''})^2,\;\;(p={\rm odd}),\\
&&A_e^{(j)}=\frac{1}{2p+1}\sum_{p'p''}^{\rm
even}(S^{(j)}_{pp'p''})^2,\;\;(p={\rm odd}),\\
&&A_c^{(j)}=\frac{1}{2p+1}\sum_{p'}^{\rm odd}\sum_{p''}^{\rm
even}(S^{(j)}_{pp'p''})^2,\;\;(p={\rm even}).
\end{eqnarray}
The summation over $p'$ and $p''$ can be performed explicitly and
$A$'s are independent of $p$.
For half-integer (integer) $j$, we obtain
\begin{eqnarray}
&&A^{(j)}_m=\frac{1}{4}(2j+3),\;\;\left(\frac{1}{4}(2j-1)\right), \\
&&A^{(j)}_e=\frac{1}{4}(2j-1),\;\;\left(\frac{1}{4}(2j+3)\right), \\
&&A^{(j)}_c=\left\{\begin{array}{ll}
0 & (j=1/2),\\
\frac{1}{4}(2j+1) & ({\rm otherwise}).
\end{array}\right.
\end{eqnarray}
One can confirm that the scaling equation for the SU($2n$) CS model is
reproduced with $J_m=J_e=J_{\rm CS}$.

From eqs. (\ref{sc1}) and (\ref{sc2}), one can identify the following
two fixed points for all $n \geq 2$:
\begin{quote}
(i) ($J_m^*,J_e^*$)=(1,1),\hspace{7mm} (ii) ($J_m^*,J_e^*$)=(1,0).
\end{quote}
Clearly, the fixed point (i) has the SU($2n$) symmetry and its local
Fermi-liquid property has been studied well.\cite{rajan}
The finite value of the coupling constant at the fixed point is an
artifact of the third-order scaling and should be replaced by infinity
in the exact scaling.

\begin{figure}
\vspace{1mm}
\begin{center}
\epsfxsize=8cm \epsfbox{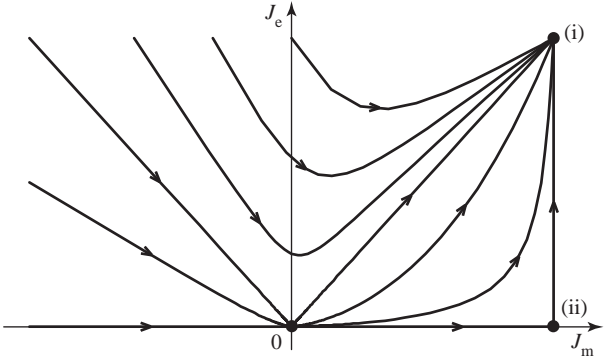}
\end{center}
\caption{The trajectory of the scaling equations for $j=5/2$ (3-fold
orbital degeneracy). There are two fixed points ($J_m^*,J_e^*$): (i)
(1,1), and (ii) (1,0).}
\label{fig1}
\end{figure}

The trajectory of the scaling equations is shown in Fig. \ref{fig1} for
the case of $j=5/2 \ (n=3)$.
Note that $J_m$ is driven to strong coupling no matter how small is
$J_e$.
On the other hand, $J_e$ can retain the zero initial value even if
$J_m$ is finite.
This aspect is related to the particle-hole symmetry in the system as
will be discussed shortly.

\begin{figure}
\vspace{1mm}
\begin{center}
\epsfxsize=8cm \epsfbox{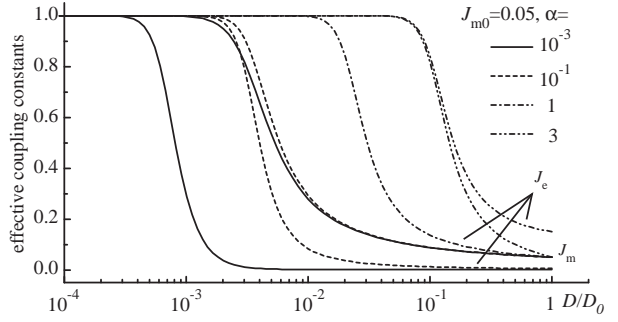}
\end{center}
\caption{The renormalization evolution for $j=5/2$.
The initial coupling constants are fixed at  $J_{m0}=0.05$ with
various values of the ratio $\alpha=J_{e0}/J_{m0}$.}
\label{fig2}
\end{figure}

For the SU($2n$) CS model with $2n=2j+1$, the characteristic energy is
given by
$T_{\rm K}^{\rm CS}=J_{\rm CS}^{1/(2n)}\exp[-1/(2nJ_{\rm CS})]$.
In the absence of $J_e$, on the other hand, the alternative energy
scale appears:
$T_{\rm K}^m=J_m^{1/(n+1)}\exp[-1/(n+1)J_m]$.
Figure \ref{fig2} shows the solution of the third-order scaling equation
for $j=5/2 \ (n=3)$.
Here the initial coupling constant is fixed at $J_{m0}=0.05$ and the
ratio $\alpha=J_e/J_m$ is varied.
It is emphasized that $J_e$ remains small  over a substantial energy
range although $J_m$ is already in the strong-coupling regime.
Hence one can identify two different characteristic energies in the
case of $\alpha\ll 1$.
The two characteristic energies merge into a single Kondo energy as
$\alpha$ approaches unity.

%
%
Now we turn to discussing the property of the new fixed point (ii)
with $J_e^*=0$.
If one performs the particle-hole transformation
$c^\dagger_{km}\to(-1)^{j-m}h_{-k-m}$,
the coupling constants $\bar{J_p}$ for the transformed system are
related to the original ones as
\begin{equation}
\bar{J_p}=(-1)^{p+1}J_p,
\end{equation}
which results from the anticommutation of fermion operators and the
parity of spherical tensors.
Thus the particle-hole symmetry is realized only for $J_p=0$ with $p$
even.
The SU($2n$) CS model with $n > 1$ breaks this symmetry since the
model necessarily has $J_e \ne 0$.

Let us discuss how the charge Kondo effect arises,
assuming that the initial coupling constants are already in the
strong-coupling regime.
The localized spin and one-particle or hole excitation from the Fermi
sea form the singlet bound state.
We represent this state as $(+,0)$ where $+$ means the binding with a
particle excitation and $0$ the singlet.
There is the particle-hole conjugate state $(-,0)$ where a hole is
bound to the localized spin.
The energies of the two states are different by the amount $J_e$.
If one reduces the magnitude of $J_e$ to zero with $J_m$ fixed in the
strong-coupling regime, the ground state with the two-fold degeneracy
in the charge degrees of freedom, $(\pm,0)$, is realized.
Therefore, at the fixed point (ii) the spin degrees of freedom is
completely compensated, but the charge degrees of freedom (isospin)
still remain.

We now construct the effective interaction around the fixed point
(ii).
In the lowest order of $t^2/J_m$, with $t$ being the hopping of
innermost conduction electrons,
there appear interaction terms which connect
the two-fold degenerate ground states.
These terms accompany
$c^\dagger_{km}h_{-k'm}$, or equivalently
$(-1)^{j-m}c^\dagger_{km}c^\dagger_{k'-m}$,
and the matrix element $\langle
+,0|c^\dagger_{km}c^\dagger_{k'-m}|-,0\rangle$ is finite.
Since the singlet state is isotropic,
the pair creation with any of $m$ has the same magnitude in the
interaction.
Thus the effective interaction around the fixed point (ii) is given by

\begin{equation}
H^*_{\rm eff}=F\sum_{m=1}^n \vec{I}_m\cdot\vec{\cal I}-h\;{\cal I}_z,
\label{iso-int}
\end{equation}
where the ``isospin'' operators for conduction electrons are defined
as
\begin{eqnarray}
&&I_m^z=\frac{1}{2}\left[\sum_{k}\left(c^\dagger_{km}c^{}_{km}+c^\dagger_{k-m}c_{k-m}\right)-1\right],\\
&&I_m^+=(-1)^{j-m}\sum_{kk'}c^\dagger_{km}c^\dagger_{k'-m},\;\;
I_m^-=[I_m^+]^\dagger,
\end{eqnarray}
and for localized electrons as
\begin{eqnarray}
&&{\cal I}_z=\frac{1}{2}\biggl[|+,0\rangle\langle
+,0|-|-,0\rangle\langle -,0|\biggr],\\
&&{\cal I}_+=|+,0\rangle\langle-,0|,\;\;
{\cal I}_-=[{\cal I}_+]^\dagger .
\end{eqnarray}
The effective interaction is equivalent to that of the multi-channel Kondo model with the overscreening.
The multiple channels come from the orbital degeneracy.
Here we have included the weak particle-hole asymmetry as the second term with $h$ in eq.(\ref{iso-int}), which is of the order of $J_e$.
Just like the multi-channel Kondo model which is unstable in the
presence of magnetic field,\cite{pang} the electric coupling $J_e$ is
the relevant perturbation in the present model.

Since the ground state has a degeneracy with $J_e=0$, the system
exhibits an anomalous behavior with the fractional entropy as in the
conventional multi-channel Kondo effect.\cite{cox}
The effective Hamiltonian is meaningful only in the case where the
renormalized coupling $J_m$ becomes strong while $J_e$ still is in the
weak-coupling regime.
In such a case only the electric response $\chi_p(\omega)=\langle
T^{(p)}_q(\omega)T^{(p)}_q\rangle$ with even $p$ shows
non-Fermi-liquid behavior.
On the contrary $\chi_p(\omega)$ with odd $p$ shows a Fermi-liquid
behavior because the spin degrees of freedom is quenched.
We are now exploring quantitatively the unique behavior of the system
by using the Wilson's numerical renormalization-group (NRG)
method, and some anomalous properties, such as the fractional entropy, have been confirmed.\cite{kusunose2}

%
%
In summary, we have investigated a multipolar exchange model with the
rotational symmetry.
Respecting the different time-reversal properties of operators, we
find a new fixed point where a non-Fermi liquid appears as a result of
the Kondo effect in the charge sector.
Although the fixed point is unstable against any perturbation breaking
the particle-hole symmetry,
the non-Fermi-liquid behavior may be observed over a substantial
energy range in the case of $J_e\ll J_m$.
Thus, more quantitative investigation of the model is highly desired.

%
%
\section*{Acknowledgements}
This work was supported by Grand-in-Aid for encouragement of Young
Scientists from the Ministry of Education, Science and Culture of
Japan and also by CREST from the Japan Science and Technology
Corporation.


\begin{thebibliography}{99}
\bibitem{multi} In leading to the usual multi-channel Kondo exchange interaction, the coupling with different $j$ partial waves is essential.
\end{thebibliography}
\end{document}